\documentclass[sigconf,nonacm]{acmart}

\usepackage{tikz}
\usetikzlibrary{shapes.geometric, arrows.meta, positioning}

\AtBeginDocument{%
  }

\renewcommand\footnotetextcopyrightpermission[1]{}
\setcopyright{none}

\begin{document}


\title{Toward Quantum-Safe Software Engineering: A Vision for Post-Quantum Cryptography Migration}
\author{Lei Zhang}
\email{leizhang@umbc.edu}
\orcid{0000-0001-9343-3654}
\affiliation{%
  \institution{University of Maryland, Baltimore County}
  \city{Baltimore}
  \state{Maryland}
  \country{USA}
}

\begin{abstract}

The quantum threat to cybersecurity has accelerated the standardization of Post-Quantum Cryptography (PQC). Migrating legacy software to these quantum-safe algorithms is not a simple library swap, but a new software engineering challenge: existing vulnerability detection, refactoring, and testing tools are not designed for PQC's probabilistic behavior, side-channel sensitivity, and complex performance trade-offs. To address these challenges, this paper outlines a vision for a new class of tools and introduces the Automated Quantum-safe Adaptation (AQuA) framework, with a three-pillar agenda for PQC-aware detection, semantic refactoring, and hybrid verification, thereby motivating Quantum-Safe Software Engineering (QSSE) as a distinct research direction.

\end{abstract}


\maketitle

\section{Introduction}

With the finalization of initial Post-Quantum Cryptography (PQC) standards from the U.S. National Institute of Standards and Technology (NIST) in 2024~\cite{BibEntry2025Jul1}, the quantum threat to cybersecurity has become a concrete software engineering problem~\cite{nather2024migrating}. The global software ecosystem must now undergo a quantum-safe cryptographic transition at massive scale~\cite{wiesmaier2021pqc}, so treating this as a routine ``library upgrade'' is dangerously simplistic.

The PQC migration represents a distinct class of software evolution challenge because PQC algorithms impose constraints that fall outside traditional maintenance paradigms. Consider the seemingly simple task of refactoring a call from RSA to a PQC signature scheme such as ML-DSA (Dilithium)~\cite{BibEntry2025Jul2}. A standard refactoring engine might correctly replace API calls, yet it would be oblivious to several non-functional but security-critical requirements: PQC signatures and keys are often orders of magnitude larger, violating latency, bandwidth, and storage assumptions; some implementations admit small but non-zero probabilistic failure rates, requiring new error-handling logic; and implementations must execute in (approximate) constant time to resist side-channel attacks~\cite{nejatollahi2019post}, so a transformation that is functionally correct can still be insecure if it alters timing behavior.

Existing software engineering tools for analysis, refactoring, and testing are not equipped to handle these semantic differences. Migration roadmaps and policy documents emphasize ``crypto-agility'' and ``PQC readiness,'' but organizations report that their progress is blocked by the lack of scalable tools for cryptographic discovery, code-level refactoring, and PQC-tailored verification~\cite{ott2019identifying,le2025enterprises, ahmed2025survey}. The PQC migration is thus hindered by a gap between high-level guidance and the software-engineering support needed to implement it in real systems.

This paper presents a vision to help close this gap. It defines \emph{Quantum-Safe Software Engineering (QSSE)} as a distinct subarea of software engineering that sits on top of existing PQC migration roadmaps and Cryptography Bill of Materials (CBOM) standards. QSSE focuses on automating cryptographic discovery, code refactoring, and continuous verification for PQC migration across large, evolving codebases and Continuous Integration/Continuous Deployment (CI/CD) pipelines, rather than on designing new cryptographic primitives or policies. Compared to classical software engineering, QSSE treats cryptographic constraints as first-class drivers of architecture, evolution, and operations, rather than as low-level library details. Building on this perspective, this paper then introduces the \emph{Automated Quantum-safe Adaptation (AQuA)} framework, a three-pillar research agenda for (P1) PQC-aware detection, (P2) semantic refactoring, and (P3) hybrid verification. 

\section{Landscape and Gaps in PQC Migration}

The broader PQC ecosystem already includes migration roadmaps such as the 7E framework and the PMMP~\cite{zhang2020quantum,zhang2023making,von2024pmmp}, as well as national guidelines and white papers from governments and industry~\cite{BibEntry2025Sep,wiesmaier2021pqc}. These efforts provide valuable high-level phases (e.g., inventory, evaluate, experiment, evolve) and emphasize cryptographic inventory, crypto-agility, and governance. 
However, from a QSSE perspective, this landscape exposes three persistent gaps. \textbf{G1: Lack of code-level semantics.} Current CBOM-oriented approaches provide a descriptive inventory of cryptographic assets but treat code largely as an opaque container; they rarely capture how cryptographic operations are embedded in control and data flow or how changes to algorithms and parameters propagate through a system~\cite{ott2019identifying,wiesmaier2021pqc}. 
\textbf{G2: Lack of refactoring patterns and automation.} Existing roadmaps and tools say what needs to be migrated and by when, but not how to systematically transform code and architectures~\cite{zhang2023making,von2024pmmp}; there is no systematic collection of PQC-aware refactoring patterns or automated engines for applying them at scale, so industrial case studies often rely on bespoke scripts. 
\textbf{G3: Lack of PQC-tailored continuous verification.} DevSecOps integrations around CBOM focus on policy compliance~\cite{ott2019identifying,wiesmaier2021pqc}, but they do not provide continuous, PQC-specific regression and non-functional checks that compare classical and post-quantum variants or assess the impact of larger keys and ciphertexts.

\section{The AQuA Framework}

\begin{figure*}[thp!]
\centering
\resizebox{\textwidth}{!}{
\begin{tikzpicture}[
    node distance=0.5cm and 1.5cm,
    auto,
    block/.style={
        rectangle, 
        draw, 
        fill=blue!10, 
        text width=3.2cm, 
        minimum height=2.2cm, 
        align=center,
        rounded corners
    },
    io/.style={
        trapezium, 
        trapezium left angle=70, 
        trapezium right angle=110, 
        draw, 
        fill=gray!20,
        text width=2.5cm,
        minimum height=1.5cm,
        align=center
    },
    line/.style={
        draw, 
        -Latex,
        thick
    }
]
\node[io] (input) {Legacy Codebase \\ \textit{\footnotesize (Source, Binaries, Configs)}};

\node[block, right=of input] (detect) {
    \textbf{Pillar I:} \\ PQC-Aware Detection
    \vspace{2mm}
    \hrule
    \vspace{2mm}
    \footnotesize Static \& Dynamic Analysis \\ ML-Assisted Scanning
};

\node[block, right=of detect] (refactor) {
    \textbf{Pillar II:} \\ Semantic Crypto-Refactoring
    \vspace{2mm}
    \hrule
    \vspace{2mm}
    \footnotesize Architectural Patterns \\ Hybrid Code Generation
};

\node[block, right=of refactor] (verify) {
    \textbf{Pillar III:} \\ Hybrid Correctness Verification
    \vspace{2mm}
    \hrule
    \vspace{2mm}
    \footnotesize Side-Channel Analysis \\ Differential Testing
};

\node[io, right=of verify] (output) {Verified, Quantum-Safe Software};

\path[line] (input) -- (detect);
\path[line] (detect) -- node[above, font=\footnotesize] {CBOM} (refactor);
\path[line] (refactor) -- node[above, font=\footnotesize] {Refactored Code} (verify);
\path[line] (verify) -- (output);

\end{tikzpicture}
}
\caption{The AQuA Framework: An integrated pipeline for PQC migration, consisting of PQC-aware detection, semantic crypto-refactoring, and hybrid correctness verification.}
\label{fig:aqua}
\end{figure*}

To address these gaps, this paper proposes AQuA, a framework for a new class of tools with PQC-aware semantic and non-functional constraints, which consists of three interdependent pillars, as illustrated in Figure~\ref{fig:aqua}.

\textbf{P1: PQC-Aware Detection. }
Existing CBOM-oriented tools already take an important first step toward PQC migration by discovering cryptographic assets and emitting inventories at the level of algorithms, libraries, and endpoints. However, from a software-engineering perspective, these inventories remain largely semantic-free: they rarely capture where cryptographic operations sit in the control and data flow of a system, which protocol roles they implement, or how changes to algorithms and parameters would propagate through surrounding code and data structures. This gap (G1) makes it difficult to move from ``we know what cryptography we use'' to ``we know how to safely transform this code.''

Pillar P1 envisions PQC-aware detection engines that enrich CBOM-style artifacts with code-level information: linking cryptographic uses to call graphs, dataflow slices, and protocol roles, and tracing transitive dependencies in the software supply chain. Realizing this vision requires new static and dynamic analyses, potentially augmented by machine learning, that can construct such enriched CBOMs in a scalable, language-agnostic way.

\textbf{P2: Semantic Crypto-Refactoring. }
Current PQC migration guidance typically reduces code-level change to phrases such as ``replace vulnerable algorithms'' or ``enable hybrid key exchange,'' leaving the concrete transformations to individual teams and ad-hoc scripts. There is no systematic collection of PQC-aware refactoring patterns that, for example, show how to restructure APIs whose message formats are affected by larger signatures, how to introduce hybrid handshakes while preserving backwards compatibility, or how to separate cryptographic concerns in tightly coupled legacy systems. This gap (G2) forces organizations to re-discover similar migration strategies in isolation and hinders automation.

Pillar P2 positions PQC-aware semantic refactoring patterns as first-class software artifacts: systematic, reusable transformation schemata that capture the preconditions, edits, and consequences of common migration scenarios, and that future tools can apply automatically or semi-automatically at scale. For instance, a pattern for migrating from RSA to ML-DSA would not only replace API calls, but also propagate size and type changes through data structures, database schemas, and protocol messages, introduce hybrid modes where required, and refactor towards crypto-agile abstractions rather than hard-coded algorithms.

\textbf{P3: Hybrid Correctness Verification. }
PQC roadmaps frequently call for verification, but existing practice tends to treat PQC components as black-box libraries: once unit tests pass and performance is acceptable, there is little support for continuous, PQC-specific regression across system evolution. Current CI/CD integrations around CBOM focus on policy compliance, not on end-to-end checks of functional and non-functional behavior after migration. This gap (G3) leaves long-term quantum-safe correctness and performance largely unmonitored.

Pillar P3 proposes hybrid correctness verification workflows that combine traditional regression testing, metamorphic and differential testing between classical and PQC variants of the same system, and selective static and dynamic analyses for side-channel resilience (e.g., approximate constant-time behavior). This work proposes integrating these checks into CI/CD pipelines as recurring, automation-friendly stages, transforming ``quantum readiness'' from a one-time audit into a continuous assurance process.

\section{Conclusion}

The transition to a quantum-safe world is no longer a purely cryptographic problem; it is a major software evolution challenge. This paper has outlined QSSE as a distinct research agenda and introduced the AQuA framework as an architectural vision for tools that support PQC-aware detection, refactoring, and verification. The goal is to spark collaboration on building and evaluating such tools so that PQC migration can be carried out safely and at scale.

\section*{Acknowledgment}

This work is supported by the NSF Grant No. 2347249.

\bibliographystyle{ACM-Reference-Format}
\bibliography{main}

@article{zhang2020quantum,
  title={Quantum advantage and the {Y2K} bug: A comparison},
  author={Zhang, Lei and Miranskyy, Andriy and Rjaibi, Walid},
  journal={IEEE Software},
  volume={38},
  number={2},
  pages={80--87},
  year={2020},
  publisher={IEEE}
}

@article{zhang2023making,
  title={Making existing software quantum safe: A case study on {IBM Db2}},
  author={Zhang, Lei and Miranskyy, Andriy and Rjaibi, Walid and Stager, Greg and Gray, Michael and Peck, John},
  journal={Information and Software Technology},
  volume={161},
  pages={107249},
  year={2023},
  publisher={Elsevier}
}

@misc{BibEntry2025Jul1,
	title = {{Workshops and timeline --- post-quantum cryptography}},
	author = {NIST Computer Security Resource Center},
	year = {2025},
	month = Jul,
	note1 = {[Online; accessed 24. Sep. 2025]},
	url = {https://csrc.nist.gov/projects/post-quantum-cryptography/workshops-and-timeline}
}

@misc{BibEntry2025Jul2,
	title = {{Selected algorithms --- post-quantum cryptography}},
	author = {NIST Computer Security Resource Center},
	year = {2025},
	month = Jul,
	note1 = {[Online; accessed 24. Sep. 2025]},
	url = {https://csrc.nist.gov/Projects/post-quantum-cryptography/selected-algorithms}
}

@misc{BibEntry2025Sep,
	title = {{NSA} releases future quantum-resistant (QR) algorithm requirements for national security systems},
	author = {National Security Agency/Central Security Service},
	year = {2025},
	month = Sep,
	note1 = {[Online; accessed 24. Sep. 2025]},
	url = {https://www.nsa.gov/Press-Room/News-Highlights/Article/Article/3148990/nsa-releases-future-quantum-resistant-qr-algorithm-requirements-for-national-se}
}

@misc{ott2019identifying,
  title={Identifying research challenges in post quantum cryptography migration and cryptographic agility},
  author={Ott, David and Peikert, Christopher and others},
  year={2019},
  eprint={1909.07353},
  archivePrefix={arXiv},
  primaryClass={cs.CY},
  url1={https://arxiv.org/abs/1909.07353}, 
}

@inproceedings{von2024pmmp,
  title={{PMMP-PQC} migration management process},
  author={Von Nethen, Nils and Wiesmaier, Alexander and Alnahawi, Nouri and Henrich, Johanna},
  booktitle={Proceedings of the 2024 European Interdisciplinary Cybersecurity Conference},
  pages={144--154},
  publisher = {ACM},
  address = {New York, NY, USA},
  year={2024}
}

@misc{le2025enterprises,
      title={Are enterprises ready for quantum-safe cybersecurity?}, 
      author={Tran Duc Le and Phuc Hao Do and Truong Duy Dinh and Van Dai Pham},
      year={2025},
      eprint={2509.01731},
      archivePrefix={arXiv},
      primaryClass={cs.CR},
      url1={https://arxiv.org/abs/2509.01731}, 
}

@misc{wiesmaier2021pqc,
  title={On {PQC} migration and crypto-agility},
  author={Wiesmaier, Alexander and Alnahawi, Nouri and Grasmeyer, Tobias and Gei{\ss}ler, Julian and Zeier, Alexander and Bauspie{\ss}, Pia and Heinemann, Andreas},
  year={2021},
  eprint={2106.09599},
  archivePrefix={arXiv},
  primaryClass={cs.CR},
  url1={https://arxiv.org/abs/2106.09599}, 
}

@inproceedings{ahmed2025survey,
  title={A survey of post-quantum cryptography support in cryptographic libraries},
  author={Ahmed, Nadeem and Zhang, Lei and Gangopadhyay, Aryya},
  booktitle={2025 IEEE International Conference on Quantum Computing and Engineering (QCE)},
  volume={1},
  pages={906--917},
  year={2025},
  organization={IEEE}
}

@article{nejatollahi2019post,
  title={Post-quantum lattice-based cryptography implementations: A survey},
  author={Nejatollahi, Hamid and Dutt, Nikil and Ray, Sandip and Regazzoni, Francesco and Banerjee, Indranil and Cammarota, Rosario},
  journal={ACM Computing Surveys (CSUR)},
  volume={51},
  number={6},
  pages={1--41},
  year={2019},
  publisher={ACM New York, NY, USA}
}

@article{nather2024migrating,
  title={Migrating software systems towards post-quantum cryptography--a systematic literature review},
  author={N{\"a}ther, Christian and Herzinger, Daniel and Gazdag, Stefan-Lukas and Stegh{\"o}fer, Jan-Philipp and Daum, Simon and Loebenberger, Daniel},
  journal={IEEE Access},
  year={2024},
  volume={12},
  number={},
  pages={132107-132126},
  publisher={IEEE}
}

\end{document}